# Sharing Linkable Learning Objects with the use of Metadata and a Taxonomy Assistant for Categorization

Valentina Franzoni[1,2], Sergio Tasso[1], Simonetta Pallottelli[1], Damiano Perri[1]

[1] Department of Mathematics and Computer Science, University of Perugia
via Vanvitelli, 1, I-06123 Perugia, Italy
{sergio.tasso, simonetta.pallottelli, valentina.franzoni}@unipg.it
[2] Department of Computer, Control, and Management Engineering "Antonio Ruberti",
Sapienza University of Rome
via Ariosto, 25, 00185 Roma, Italy

**Abstract.** In this work, a re-design of the *Moodledata* module functionalities is presented to share learning objects between e-learning content platforms, e.g., Moodle and G-Lorep, in a linkable object format. The e-learning courses content of the Drupal-based Content Management System G-Lorep for academic learning is exchanged designing an object incorporating metadata to support the reuse and the classification in its context. In such an Artificial Intelligence environment, the exchange of Linkable Learning Objects can be used for dialogue between Learning Systems to obtain information, especially with the use of semantic or structural similarity measures to enhance the existent Taxonomy Assistant for advanced automated classification.

## 1 Introduction

In recent years, e-learning systems became part of the Artificial Intelligence revolution, [1] where machines and intelligent systems can interchange information through an Internet connection. Before this concept, the Linked Data Networks already connected information objects, where the World Wide Web became the path through the creation of a global information space comprising linked documents. [2–4]

G-Lorep (Grid Learning Object Repository) [5, 6] is a federation of distributed repositories containing Learning Objects (LO)s from various academic sources, developed as a Learning Managing System (LMS) at the *Department of Mathematics and Computer Science* of the *University of Perugia*, where it can boast an excellent operativity of about twenty years. The implementation of G-Lorep involved the *Department of Chemistry, Biology, and Biotechnologies* of the University of Perugia, as well as other European universities, e.g., Genoa and Thessaloniki. G-Lorep is also useful to support the student for the preparation to tests, including specialized ones such as the EChemTest® e-test, providing increasingly improved LOs, thanks to the cooperative effort of the multidisciplinary team of management members. [8]

Regarding sharing LO Metadata (LOM), G-Lorep leverages different Moodle [7] modules, e.g. *Moodledata*, [8] to share a LOM between a Content Management System (CMS) and a Learning Management System, such as G-Lorep (managed by Drupal) and Moodle, where the same design can be applied with different modules to any other CMS and LMS.

## 2 The G-LOREP-MOODLE connections

In previous works [3, 10-11] we integrated the *Moodledata* module into Drupal, allowing the platform manager (i.e., the user with administrator permissions) to search and download files on the Moodle server, and to upload and classify them on Drupal as LOMs.

Regarding the continuous update of Drupal, we achieved two main objectives:

a) adapt the module to the new versions of Drupal;
b) extend the module by adding new features:

- Download files from Moodle.
- Create a new Linkable Learning Object (LLO) from the downloaded data.
- Upload the LLO to Drupal in G-Lorep (see Fig.1).

The module customization, for the *Moodledata* upload, mainly includes three files with extensions "**.module**", "**.install**", "**.info**".

A *README.txt* file provides a brief installation guide for the module, integrated by the online help.

## 2.1 Design and Environment Setting

As a first step, some features must be set to use the *Moodledata* module. The Drupal side sees two necessary modules for *moodledata* to be installed and set:

- *Collabrep*, to manage the information transmission in the G-Lorep federation.
- *Linkable Object* (LO), to manage the node and check user permissions, where a LO can be defined as a *Learning Object* or a *Software Attachment*.

Such modules can be set in the *Modules* section of Drupal *Administration* settings.

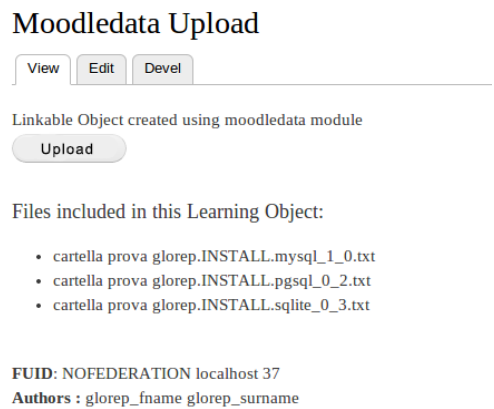

*Fig. 1*: The Upload-Creation-Download

## 2.2 Setting the Drupal web-service

As in the case of download from Moodle to Drupal, [8] also for the case of downloading from Drupal files to be uploaded in Moodle, we need to manage a web service in Drupal. The *Moodledata* menu is used to access the module's features, which contains a short description of the information objects to access. The first is the *Settings* invoice, to set the Moodle access, with the following fields:

- Domain *address* of the Moodle installation.
- *Username*, to access the Moodle account.
- Login *Password*.

A connection to the web-service is attempted upon successful accomplishment of such operations,, and the access token is requested and kept for all the queries of the session. The module is now available to authenticated G-Lorep users.

**2.3 G-LOREP *moodledata* download and storage**

As easily understandable, the primary functions of the module regard the appropriate managing of the file downloads from G-Lorep.
A file can be searched specifying a string including the full file name or its parts. An *Advanced Search* is available to refine a search combining additional parameters, e.g., course name, file author, and the timestamp of the last editing. The functions are:

1. **Search** (clicking on the *Search button)*: files matching the search criteria set are included in a table with the following fields:
   - *select item*: a checkbox for file selection;
   - *course name*: the course containing the file;
   - *filename*: the file name.
2. **Download** (*Download selected items* feature): the selected files are downloaded and stored in a temporary cache in G-Lorep. This intermediate step between download and upload is transparent to the user, for better management of the process in order to prevent and avoid human errors.
3. **Metadata addition**: the most exciting part of this work is the critical update and re-design of the module, replacing the database connection with the use of Moodle web-services to include a metadata exchange. Such semantic addition is to facilitate information retrieval of LLOs, exploited adding automatically content and context-related metadata using the G-Lorep taxonomy. LMSs like Moodle allow to add courses to categories, such as in a CMS, and this feature perfectly fits the feature of classification metadata exploited through the G-Lorep *Taxonomy Assistant*, which will be presented in section 3.
4. **Save**: the user saves the files shown after a final check. An individual file loaded individually will maintain the name given when loading, while a collection of files will gain the name of both the folder and the file.

In addition to the file name, the other fields in the table are:

- *author*: file author;
- *format*: file type;
- *description*: a string describing the file, or folder given by the Moodle user during loading.
- *last modified*: the last editing timestamp date.

**2.4 Moodledata Upload in Moodle**

Another primary function of the module provides the possibility to upload to Moodle the Drupal files downloaded in the *moodledata* table. Fig. 2 shows the G-Lorep linkable object ready to be shared.

The evolution of G-Lorep providing the automation of the upload is a continuous work in progress. The saved LLO, for now, can be uploaded using the import function of Moodle.

It should be noted that additional metadata, e.g., the files list and the taxonomy metadata do not appear in the interface of the LLO (see Fig. 2) because it shows only the editable content. The additional metadata are attached to the Linkable Object and can be visible from the G-Lorep course page, or directly in Moodle after the importing step.

*Fig. 2*: The G-LOREP moodledata object (LLO interface)

## 3 G-LOREP Taxonomy metadata

### 3.1 Metadata storage in G-LOREP through the Drupal API Field

The realization of the LO module makes use of the innovations introduced since the 7.x version of the Drupal CMS by the *Field API*, to customize fields attachable to Drupal entities regarding storage, editing, loading, and view.
To this aim, two data structures are defined:

- *Field*: defines the format of data to attach to the entity.
- *Instance*: allows instantiating the data type.

In particular, in order to implement the LLO module, we chose the *body* and *files* Field APIs, respectively dealing with the description of the node and the information about the files attached to the LLO. The main advantage of Field APIs is the metadata management because Drupal itself takes care of the database management. For instance, about taxonomy metadata, the mentioned fields, stored in the *file_managed* table of the Drupal database, can be attached to the Linkable Object automatically.
The changes are propagated by merely creating the object with the attached metadata, which will be stored in the shared database; a future call to the server that executes the download routine will get all the metadata within the object. At the creation step, the categories are visible to the user and can be selected. An assistant helps the user to select the best category (see fig. 3). The same happens for the synonyms used for tagging the object. [10–12]

### 3.2 The G-LOREP Taxonomy Assistant

The classification of educational objects (e.g., LLOs) is contributory to the organization of knowledge units in the realm context, allowing efficient reuse and information retrieval through automatic labeling and management. This enhancement is based on appropriate tree classifications (e.g., hierarchies, taxonomies) or graph classifications (e.g., ontologies) linking related entities. Recently, several proposals to manage semantic information through web-based queries shown impressive performances with fewer problems than ontologies. [11-12, 16-29]
In our work in G-Lorep, we adopted the *Drupal Taxonomy Assistant* (TA), [9] where future works will include web-based similarity measures. TA is a module able to interact with the LLO, via the *Linkable Object* and the *dis_cat* modules. TA analyses the content of the text and the LO description and automatically proposes the best category, then selectable and editable by the user.

```
Title: *
Coupling Quantum Interpretative Techniques: Another Look at Chemical Mechanisms in Organic Reactions
Description (20 words):
A cross ELF/NCI analysis is tested over prototypical organic reactions. The synergetic use of ELF and NCI enables
the understanding of reaction mechanisms since each method can respectively identify regions of strong and
weak electron pairing. Chemically intuitive results are recovered and enriched by the identification of new
features. Noncovalent interactions are found to foresee the evolution of the reaction from the initial steps. Within
NCI, no topological catastrophe is observed as changes are continuous to such an extent that future reaction
steps can be predicted from the evolution of the initial NCI critical points. Indeed, strong convergences through
the reaction paths between ELF and NCI critical points enable identification of key interactions at the origin of the
bond formation. VMD scripts enabling the automatic generation of movies depicting the cross NCI/ELF analysis
along a reaction path (or following a Born–Oppenheimer molecular dynamics trajectory) are provided as
Supporting Information.

Categories suggested by Taxononomy assistant:
This is the list of categories that are compatible with the text and their value inherence (Hin value) and relevance (max for single term% | total %):
(Remember that you haven't yet selected a category from the vocabularies)
541.2 - Theoretical Chemistry (keywords:'reaction' 'molecular bond' 'quantum' )(Hin Value: 100) Relevance: (max:2.9%) | (Tot:15.9%)
541.36 - Thermochemistry & Thermodynamics (keywords:'reaction' 'formation' 'point' )(Hin Value: 35.3) Relevance: (max:1.4%) | (Tot:7.2%)
541.39 - Chemical reactions (keywords:'reaction' )(Hin Value: 25.5) Relevance: (max:1.4%) | (Tot:5.8%)
515.78 - Special topics of functional analysis (keywords:'analysis' )(Hin Value: 17.6) Relevance: (max:1.4%) | (Tot:5.8%)
515.73 - Topological vector spaces (keywords:'topological' 'continuous' )(Hin Value: 11.8) Relevance: (max:1.4%) | (Tot:2.9%)
541.34 - Solutions Chemistry (keywords:'point' )(Hin Value: 9.8) Relevance: (max:1.4%) | (Tot:2.9%)
543.6 - Non-Optical Spectroscopy (keywords:'electron' 'analysis' )(Hin Value: 9.8) Relevance: (max:1.4%) | (Tot:4.3%)
514.7 - Analytic Topology (keywords:'analysis' )(Hin Value: 7.8) Relevance: (max:1.4%) | (Tot:1.4%)
547.2 - Organic Chemical Reactions (keywords:'reaction' )(Hin Value: 7.8) Relevance: (max:1.4%) | (Tot:1.4%)
543.2 - Classical Methods (keywords:'analysis' )(Hin Value: 7.8) Relevance: (max:1.4%) | (Tot:2.9%)
512.5 - Linear Algebra (keywords:'topological' )(Hin Value: 6.5) Relevance: (max:1.4%) | (Tot:2.9%)
547 - Organic Chemistry (keywords:'organic' )(Hin Value: 3.9) Relevance: (max:1.4%) | (Tot:1.4%)
514.2 - Algebraic Topology (keywords:'topological' )(Hin Value: 3.9) Relevance: (max:1.4%) | (Tot:1.4%)
543.5 - Optical Spectroscopy (Spectrum Analysis) (keywords:'molecular' )(Hin Value: 3.9) Relevance: (max:1.4%) | (Tot:1.4%)
519.5 - Statistical Mathematics (keywords:'analysis' )(Hin Value: 3.9) Relevance: (max:1.4%) | (Tot:1.4%)
548.8 - Physical and Structural Crystallography (keywords:'method' )(Hin Value: 3.9) Relevance: (max:1.4%) | (Tot:1.4%)
543.8 - Chromatography (keywords:'analysis' )(Hin Value: 3.9) Relevance: (max:1.4%) | (Tot:1.4%)
515 - Analysis (keywords:'analysis' )(Hin Value: 3.9) Relevance: (max:1.4%) | (Tot:1.4%)
514.3 - Topology of Spaces (keywords:'point' )(Hin Value: 2.6) Relevance: (max:1.4%) | (Tot:1.4%)
541 - Physical Chemistry (keywords:'molecular' )(Hin Value: 2) Relevance: (max:1.4%) | (Tot:1.4%)
518 - Numerical Analysis (keywords:'method' )(Hin Value: 1.3) Relevance: (max:1.4%) | (Tot:1.4%)
Depending on your choices and the text entered in the fields 'title' and 'description' Taxonomy Assistant suggests the categories that are most relevant to them.
```

*Fig.3*: *User input and categories suggested by the TA*

The TA algorithm is based on pattern-matching between keywords and the terms of its thesaurus. The editability of the suggested category might introduce human typing mistakes, but its excellent productivity feature is to allow a fast creation of a brand-new thesaurus whenever needed, e.g., when the actual meaning of a word appears in different places with different connotations, and some additions could be implemented in the classification phase. Moreover, in this way the updating phase of the thesaurus is much more comfortable, maintaining it updated as it is used.

In the G-Lorep TA, categories are organized as a forest graph, where each tree epitomizes a science area (e.g., Computer Science, Mathematics, Physics, Chemistry, Biology) filled with sub-categories: the most detailed appear as tree leaves and the most general ones as roots. Such a scheme is compliant with various classification structures, in particular with the *Dewey Decimal Classification* (DDC), [13, 14], which G-LOREP flowchart is sketched in Fig. 4. Out of the input information, the TA builds the classification step applying the following text filtering:

1. lower case conversion;
2. deletion of non-essential characters (e.g., punctuation marks, parentheses, apostrophes);
3. check and deletion of *stopwords*, i.e., words of slight interest, usually not considered particularly meaningful by search engines [15, 16] (e.g., name articles, pronouns, adverbs, and other words whose interest is lowered by frequent usage).

The goal of this filtering process is to obtain a set of keywords, deprived of superfluous terms, for querying the database. [9] In order to increase the likelihood of significant correspondences, a stemming mechanism based on standard syntax guidelines (e.g., for singular/plural extensions) enriches the TA. Each category is then associated with a thesaurus set of terms: synonyms are considered not only as single words but also as sets because they can include expressions. [20]

The fraction given by the number $X$ of keywords found in the description of the LO and the total number $N$ of keywords defining the synonym is called *coverage*. [21–23] Coverage is used to compare the user keywords with synonyms sets contained in the taxonomy. For the records obtained from the query, each synonym is also split to subclasses of defined length windows, whose presence in the database is finally counted. The count is taken as a measure of the *inherence*, considered as the **similarity between the synonyms and the user text**, [23–26] expressing the adequacy of a category assignment in the Learning Object context. [27–30]

### 3.3 Synonym inherence as an effectiveness index

The inherence of a synonym $H_s$ with a LO is the extent of pertinence $P$ of the agreed synonym in the description of a LO. Hence, $H_s$ can be taken as an effectiveness measure, with some assumptions:

a) let the value of $H_s$ be as high as the number $N$ of words composing the synonym;
b) let the value of $H_s$ for partial coverage depending on the recurrence of a word among those composing the synonym (a coverage of "1 out of 3" must result in a $H_s$ value higher than the one associated with coverage of "1 out of 4").

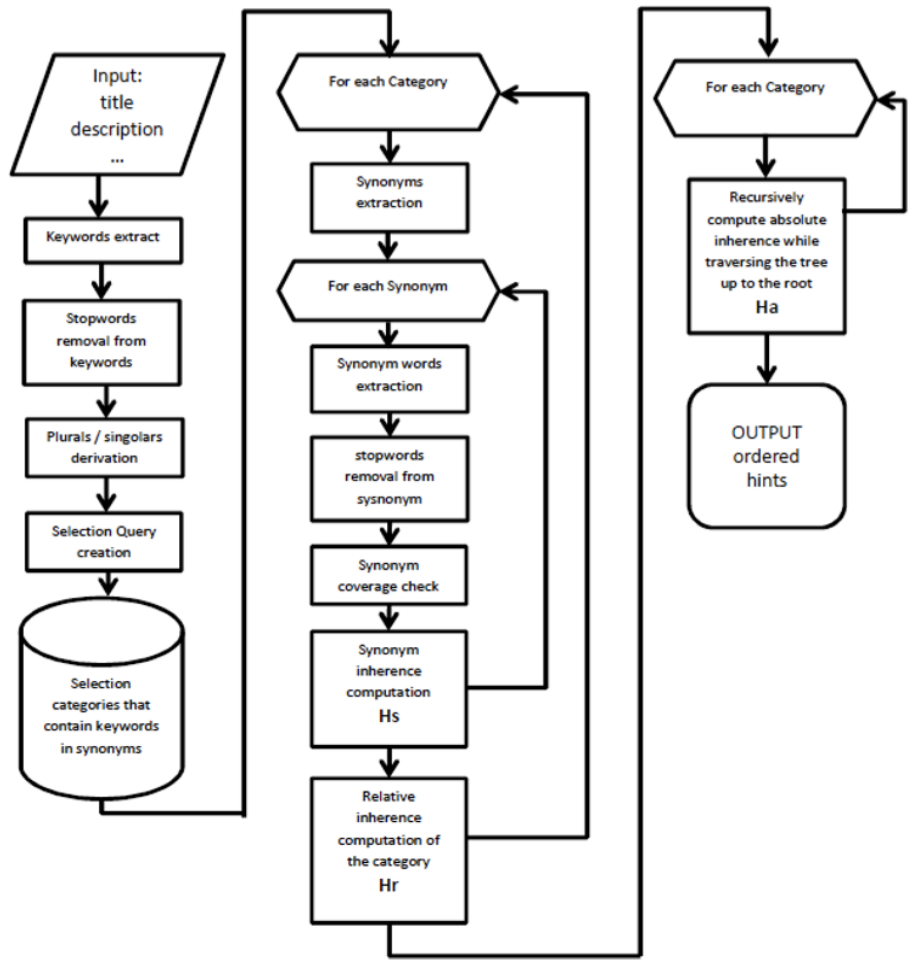

*Fig. 4: The TA algorithm scheme*

To give a precise algebraic formulation of $H_s$ we give the following definitions:

- $R_i$ is the number of occurrences of the words in synonym $i$;
- $S_i$ is the number of words composing synonym $i$;
- $K$ is the total number of valid keywords;

- $U_i = R_i/S_i$ is the ratio between the number of occurrences of the words $R_i$ found in the synonym $i$ and the number of words $S_i$ found in it;
- $P_i = S_i/K$ is the ratio between the words observed in the synonym $i$ and the total number of considered keywords $K$ (that is called either pertinence or relevance).

The simple formulation of $H_s$ was initially taken to be the power-like formula:

$$H_r = \sum_{i=1}^{\#of\,Synonyms} R_i^{U_i} U_i P_i \Rightarrow H_r = \frac{1}{K} \sum_{i=1}^{\#of\,Synonyms} \left( R_i^{\left(\frac{R_i}{S_i}\right)} \frac{R_i^2}{S_i} \right) \quad (1)$$

where the result of (1) will be *1* if there is a "*1 out of N*" coverage, and *N* for a complete coverage "*N out of N*". The $H_s$ value is the same in both cases. Since such value is in conflict with the b) requirement, its behaviour can be fixed multiplying the $H_s$ function by $U_i$:

$$H_s = R_i^{\left(\frac{R_i}{S_i}\right)} \frac{R_i}{S_i} \quad (2)$$

### 3.4 Inherence of categories

The meaning of repeated partial occurrences in synonyms within a category can be calibrated to obtain the most appropriate efficacy measures:

- *Relative inherence of a category* ($H_r$): the sum of all the synonym inherences $H_s$ of the category.
- *Absolute inherence of a category* ($H_a$): the sum of the relative inherences $H_r$ associated with all the categories in the root/node path, weighted by the related level *d*.

The relative inherence $H_r$ can be expressed as:

$$H_r = \sum_{i=1}^{\#of\,synonyms} H_{s_i} \quad (4)$$

Expanding the formula through the substitution of $H_s$ we obtain:

$$H_r = \sum_{i=1}^{\#of\,Synonyms} R_i^{U_i} U_i P_i \Rightarrow H_r = \frac{1}{K} \sum_{i=1}^{\#of\,Synonyms} \left( R_i^{\left(\frac{R_i}{S_i}\right)} \frac{R_i^2}{S_i} \right) \tag{5}$$

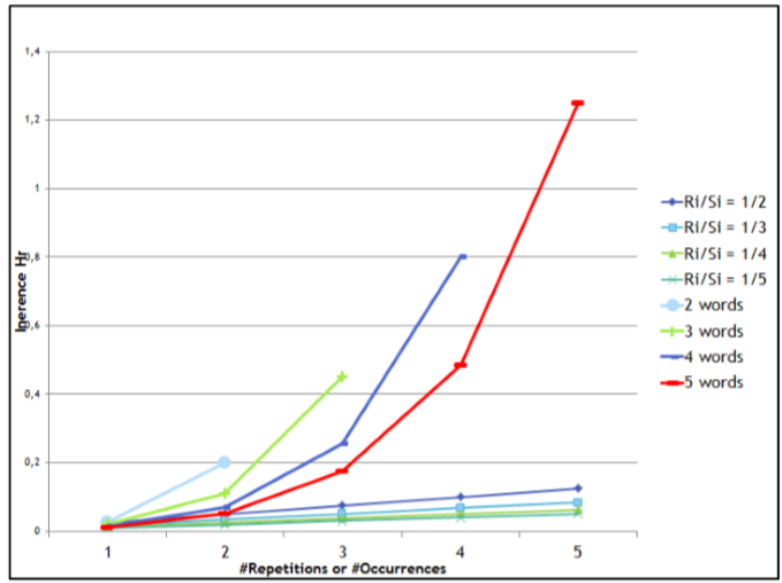

*Fig. 5: The $H_r$ values plotted as a function of the number # of repetitions (or occurrences) for different numbers of words composing the synonym*

To obtain an algebraic construction of $H_a$ we express the following positions:

- $d$ is the depth of the node,
- $R_{i,d}$ is the number of matching words in the synonymous $i$ at level $d$,
- $S_{i,d}$ is the number of words constituting synonym $i$ at level $d$,
- $U_{i,d} = R_{i,d}/S_{i,d}$ is the ratio between the number of occurrences and the number of words constituting the synonymous $i$ at level $d$.

We can, finally, work out the absolute inherence of a category obtained by adding the relative inherence value of each ancestor category *i* lying on the path from the root to the considered category multiplied by the level *d* to which the category belongs (see Fig.5):

$$H_a = \sum_{d=1}^{depth} H_{r_d} d \tag{6}$$

## 4 Searching LO with a semantics-based AI

Recently, Artificial Intelligence (AI) has been introduced to e-learning systems. In our environment, the exchange of Linkable Learning Objects can be used for dialogue between Learning Systems to obtain information, especially with the use of semantic or structural similarity measures to enhance the existent Taxonomy Assistant for advanced automated classification.

The G-LOREP platform and Moodle, being open-source, are particularly suitable to be enhanced with AI-based tools, e.g., an intelligent assistant to support information exchange. A feasible solution is to develop an assistant based on existing devices, e.g., Google Home, Amazon Alexia, Microsoft Cortana, which offer a set of APIs for voice control.

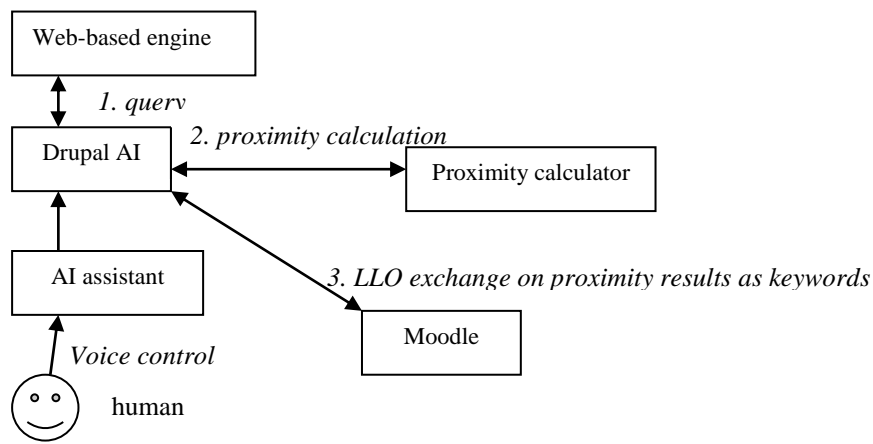

*Fig. 6 AI-based LLO exchange*

In such a web-based environment, web-based similarity measures, which rely on search engines (e.g., Google, Bing) to obtain frequency values used for semantic proximity calculation, [27, 31, 32] can be easily integrated into an AI to obtain a set of LLOs to exchange among learning systems. Fig.6 depicts the scheme of an AI-based LLO search and exchange.

## 5 Conclusions and future work

Updating an integrated module on Drupal allows downloading files managed within the G-LOREP federation and share them in the form of Linkable Learning Objects to Moodle or a Linked Data network. The future development of the module will involve a plug-in to automate the procedures for activating the *moodledata* service on the Moodle side.

We also described the use of the Taxonomy assistant (TA) automatically proposing a set of classification elements to the user in the delicate procedure of cataloging an LLO. TA proposes to the user the correct place for a LO in the taxonomy tree. Future activities will address the use of the taxonomy to link the Linkable Object for a Linked Data network. Another improvement will come from the integration in the G-Lorep TA of web-based similarity calculations, such as the *PMING distance* [31, 33] or of other structural or semantic similarity [32] measures to improve the taxonomy prediction [34–36] or the term selection, as proposed and partially explained in [37].

In such an Artificial Intelligence environment, the exchange of Linkable Learning Objects can be used for dialogue between Learning Systems or between a Learning AI and a human user for advanced human-machine communication. [29, 30, 31]

## Acknowledgments

Thanks are due to EGI and IGI and the related COMPCHEM VO for the use of Grid resources. The authors also thank the authors of previous works of the project, cited in this paper.

## Reference

1. Franzoni, V., Milani, A., Nardi, D., Vallverdú, J.: Emotional machines: The next revolution. Web Intell. 17, 1–7 (2019).
2. Noia, T. Di, Ostuni, V.C., Tomeo, P., Sciascio, E. Di: SPrank: Semantic Path-

Based Ranking for Top-N Recommendations Using Linked Open Data. ACM Trans. Intell. Syst. Technol. 8, 9:1--9:34 (2016).
3. Bizer, C., Heath, T., Berners-Lee, T.: Linked Data - The Story So Far. Int. J. Semant. Web Inf. Syst. (2010).
4. Heath, T., Bizer, C.: Linked Data: Evolving the Web into a Global Data Space. Synth. Lect. Semant. Web Theory Technol. (2011).
5. Pallottelli, S., Tasso, S., Pannacci, N., Costantini, A., Lago, N.F.: Distributed and collaborative learning objects repositories on Grid networks. In: Lecture Notes in Computer Science (including subseries Lecture Notes in Artificial Intelligence and Lecture Notes in Bioinformatics) (2010).
6. Baioletti, M., Milani, A., Poggioni, V., Rossi, F.: Experimental evaluation of pheromone models in ACOPlan. Ann. Math. Artif. Intell. 62, 197–217 (2011).
7. Marcato, E., Scala, E.: Moodle. In: Handbook of Research on Didactic Strategies and Technologies for Education (2012).
8. Tasso, S., Pallottelli, S., Gervasi, O., Rui, M., Laganà, A.: Sharing learning objects between learning platforms and repositories. In: Lecture Notes in Computer Science (including subseries Lecture Notes in Artificial Intelligence and Lecture Notes in Bioinformatics) (2018).
9. Tasso, S., Pallottelli, S., Ciavi, G., Bastianini, R., Laganà, A.: An efficient taxonomy assistant for a federation of science distributed repositories: A chemistry use case. In: Lecture Notes in Computer Science (including subseries Lecture Notes in Artificial Intelligence and Lecture Notes in Bioinformatics) (2013).
10. Tasso, S., Pallottelli, S., Ferroni, M., Bastianini, R., Laganà, A.: Taxonomy management in a federation of distributed repositories: A chemistry use case. In: Lecture Notes in Computer Science (including subseries Lecture Notes in Artificial Intelligence and Lecture Notes in Bioinformatics) (2012).
11. Tasso, S., Pallottelli, S., Rui, M., Laganá, A.: Learning objects efficient handling in a federation of science distributed repositories. In: Lecture Notes in Computer Science (including subseries Lecture Notes in Artificial Intelligence and Lecture Notes in Bioinformatics) (2014).
12. Tasso, S., Pallottelli, S., Bastianini, R., Lagana, A.: Federation of distributed and collaborative repositories and its application on science learning objects. In: Lecture Notes in Computer Science (including subseries Lecture Notes in Artificial Intelligence and Lecture Notes in Bioinformatics) (2011).
13. Mitchell, J.S., Vizine-Goetz, D.: Dewey Decimal Classification (DDC). In: Encyclopedia of Library and Information Sciences, Third Edition (2016).
14. McClelland, M.: Metadata Standards for Educational Resources, (2003).
15. Franzoni, V., Milani, A., Biondi, G.: SEMO: A semantic model for emotion recognition in web objects. In: Proceedings - 2017 IEEE/WIC/ACM International Conference on Web Intelligence, WI 2017 (2017).
16. Franzoni, V., Biondi, G., Milani, A.: A web-based system for emotion vector extraction. (2017).

17. Franzoni, V., Poggioni, V.: Emotional book classification from book blurbs. In: Proceedings - 2017 IEEE/WIC/ACM International Conference on Web Intelligence, WI 2017 (2017).
18. Franzoni, V., Poggioni, V., Zollo, F.: Can we infer book classification by blurbs? In: CEUR Workshop Proceedings (2014).
19. Franzoni, V., Poggioni, V., Zollo, F.: Automated classification of book blurbs according to the emotional tags of the social network Zazie. In: CEUR Workshop Proceedings (2013).
20. Franzoni, V., Leung, C.H.C., Li, Y., Mengoni, P., Milani, A.: Set similarity measures for images based on collective knowledge. (2015).
21. Franzoni, V., Milani, A.: A pheromone-like model for semantic context extraction from collaborative networks. In: Proceedings - 2015 IEEE/WIC/ACM International Joint Conference on Web Intelligence and Intelligent Agent Technology, WI-IAT 2015 (2016).
22. Pallottelli, S., Franzoni, V., Milani, A.: Multi-path traces in semantic graphs for latent knowledge elicitation. In: ICNC. pp. 281–288. IEEE (2015).
23. Franzoni, V., Milani, A.: Heuristic semantic walk for concept chaining in collaborative networks. Int. J. Web Inf. Syst. 10, (2014).
24. Franzoni, V., Milani, A.: A semantic comparison of clustering algorithms for the evaluation of web-based similarity measures. (2016).
25. Leung, C.H.C., Li, Y., Milani, A., Franzoni, V.: Collective evolutionary concept distance based query expansion for effective web document retrieval. (2013).
26. Franzoni, V., Mencacci, M., Mengoni, P., Milani, A.: Heuristics for semantic path search in Wikipedia. (2014).
27. Franzoni, V., Milani, A., Pallottelli, S., Leung, C.H.C., Li, Y.: Context-based image semantic similarity. In: 2015 12th International Conference on Fuzzy Systems and Knowledge Discovery, FSKD 2015 (2016).
28. Franzoni, V.: Context Extraction by Multi-path Traces in Semantic Networks. In: Rr. {IEEE} Computer Society (2016).
29. Franzoni, V., Milani, A.: Semantic context extraction from collaborative networks. In: Proceedings of the 2015 IEEE 19th International Conference on Computer Supported Cooperative Work in Design, CSCWD 2015 (2015).
30. Franzoni, V., Mencacci, M., Mengoni, P., Milani, A.: Semantic heuristic search in collaborative networks: Measures and contexts. In: Proceedings - 2014 IEEE/WIC/ACM International Joint Conference on Web Intelligence and Intelligent Agent Technology - Workshops, WI-IAT 2014 (2014).
31. Franzoni, V., Milani, A.: PMING distance: A collaborative semantic proximity measure. In: Proceedings - 2012 IEEE/WIC/ACM International Conference on Intelligent Agent Technology, IAT 2012 (2012).
32. Franzoni, V., Milani, A.: Structural and semantic proximity in information networks. (2017).
33. Biondi, G., Franzoni, V., Li, Y., Milani, A.: Web-based similarity for emotion

recognition in web objects. In: Proceedings - 9th IEEE/ACM International Conference on Utility and Cloud Computing, UCC 2016 (2016).
34. Chiancone, A., Milani, A., Poggioni, V., Pallottelli, S., Madotto, A., Franzoni, V.: A multistrain bacterial model for link prediction andrea chiancone. In: ICNC. pp. 1075–1079. IEEE (2015).
35. Franzoni, V., Chiancone, A., Milani, A.: A Multistrain Bacterial Diffusion Model for Link Prediction. Int. J. Pattern Recognit. Artif. Intell. 31, (2017).
36. Chiancone, A., Franzoni, V., Li, Y., Markov, K., Milani, A.: Leveraging Zero Tail in Neighbourhood for Link Prediction. In: {WI-IAT} {(3)}. pp. 135–139. {IEEE} Computer Society (2015).
37. Franzoni, V., et al.: Set Semantic Similarity for Image Prosthetic Knowledge Exchange. In: Lecture Notes in Computer Science (including subseries Lecture Notes in Artificial Intelligence and Lecture Notes in Bioinformatics) (2019).
38. Franzoni, V., Milani, A., Vallverdú, J.: Emotional affordances in human-machine interactive planning and negotiation. In: Proceedings - 2017 IEEE/WIC/ACM International Conference on Web Intelligence, WI 2017 (2017).
39. Brackett, M.A.: The Emotion Revolution: Enhancing Social and Emotional Learning in School: Enhancing Social and Emotional Learning in School, (2016).